\documentclass[twoside,leqno,twocolumn]{article}

\usepackage{graphicx}
\usepackage{balance}
\usepackage{subfigure}
\usepackage{multirow}
\usepackage{color}
\usepackage{amsopn}
\usepackage{mathrsfs}
\usepackage{mathtools}
\usepackage{amsmath}
\usepackage{amsfonts}
\usepackage{enumerate}
\usepackage{algorithm}
\usepackage{algorithmic}

\begin{document}

\title{Integrating Topic Models and Latent Factors for Recommendation \vspace{-10pt}}

\author{
Danis J. Wilson\thanks{danisjwilson@gmail.com}\\
\and
Wei Zhang\thanks{weizhangjames@gmail.com}\\
}
\date{}

\maketitle

\begin{abstract}
Nowadays, we have large amounts of online items in various web-based applications, which makes it an important task to build effective personalized recommender systems so as to save users' efforts in information seeking. One of the most extensively and successfully used methods for personalized recommendation is the Collaborative Filtering (CF) technique, which makes recommendation based on users' historical choices as well as those of the others'. The most popular CF method, like Latent Factor Model (LFM), is to model how users evaluate items by understanding the hidden dimension or factors of their opinions. How to model these hidden factors is key to improve the performance of recommender system. In this work, we consider the problem of hotel recommendation for travel planning services by integrating the location information and the user's preference for recommendation. The intuition is that user preferences may change dynamically over different locations, thus treating the historical decisions of a user as static or universally applicable can be infeasible in real-world applications. For example, users may prefer chain brand hotels with standard configurations when traveling for business, while they may prefer unique local hotels when traveling for entertainment. In this paper, we aim to provide trip-level personalization for users in recommendation.
\end{abstract}


\section{Introduction}

Hotel recommendation is very important in various trip planning services. Hotel recommendation is particularly different from standard recommendation problems because of the unique features in the hotel domain. First, users' preferences to hotel are destination dependent and related to the goal of the trip, e.g., if a user is on business trip, then the user may be interested in chain brand hotels that they are familiar with so that the hotel service is standard and quality guaranteed and the trip can be confidently managed. However, if the users is on an entertainment trip, then the user may prefer local hotels that can show the local specialty of the destination. As a result, hotel recommendation should consider the particular goal of the trip. Second, because hotel recommendation is highly destination dependent, then user's previous interaction histories may not very correctly predict the target trip. If we blindly apply traditional standard recommendation models, then the recommendations may be dominated by the items that are similar to the user's previously taken hotels, which even may be not located in the destination location, resulting in wrong recommendations. As a result, carefully designed hotel recommendation models are highly needed for this particular domain so as to provide satisfactory recommendation performances.

If we simply recommend items with local features, it cannot lead to good recommendations. For instance, a user dose not like cheap hotel, then he or she may not try cheap hotels in anywhere. In order to solve this problem, we should consider the user preference, the local features and the willingness of the user to try new things whose features are different from their previous experiences. As a result, we need not have restricted ourselves to user's previous exposed hotels or hotel brands, rather, we can optimize the hotel space directly as whole for model learning and hotel recommendation, which gives us the Global and Local feature model for Hotel Recommendation (GloHoRec) to be introduced in this paper.

To validate the effectiveness of our model, we conduct personalized hotel recommendation based on the user ratings and textual reviews from a major review service website Tripadvisor. Based on our preliminary analysis, we integrate the user and item latent factors, which is similar to previous work, but differently, we transform the product of latent factors as a whole to connect the modeling principle of LFM and LDA in a single learning framework, so as to verify the performance against the methods that work on each separate latent factor in previous work.

The rest of the paper is organized as follows: We summarize the related work in Section \ref{sec:related}, and then present our Global and Local feature model for Hotel Recommendation (GloHoRec) in Sections \ref{sec:model}, \ref{sec:model-cold}, and \ref{sec:optimization}. We conduct case studies to support our intuition of modeling in Section \ref{sec:case}, and present detailed experimental evaluations in Section \ref{sec:experiment}. Finally, we conclude this work in Section \ref{sec:conclusion}.

\section{Related Work}
\label{sec:related}

Recommender systems are important applications in modern web-based services. Many recommendation models are proposed for making accurate predictions. One example is collaborative filtering \cite{su2009survey}, which aims to make recommendations by using interaction information from users. For example, user-based collaborative filtering makes recommendation based on similar users \cite{resnick1994grouplens}, item-based collaborative filtering makes recommendation based on item similarity \cite{sarwar2001item}, while matrix factorization-based methods makes recommendation by learning user and item latent vectors in a latent space \cite{koren2009matrix}. Some probabilistic models are also used for recommendation, such as probabilistic matrix factorization \cite{mnih2008probabilistic}. Recommendation algorithms are also used for various different applications, such as movie recommendation, product recommendation, music recommendation, job recommendation, hotel recommendation, book recommendation, etc. Since different application scenarios have different content information and specific needs, it is important to develop different algorithms tailored to different applications. One such approach is hybrid recommendation \cite{burke2002hybrid}, which can make use of different content information and also collaborative learning for recommendation. In this paper, our proposed model belongs to the hybrid recommendation category.

\section{Local and Global Features for Recommendation}\label{sec:model}

In this section, we analysis the relationship between local and global features in the scenario of personalized hotel recommendation, and exposit the crucial requirements to bridge the models for the task. As shown in figure one, we can consider the relationship between user, item and feature as a chain structure, i.e., user $U$'s preference on item $I$ is through features $K$. For example, a user may like a particular hotel because of the features \textit{service} and \textit{price} of the hotel.

\begin{figure}[t]
	\centering
	\hspace{-10pt}
	\includegraphics*[viewport=12mm 14mm 89mm 31mm, width=0.6\linewidth,height=0.05\textheight]{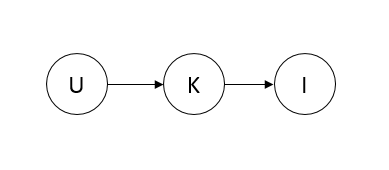}
	\caption{The dependencies of users, items and features in recommendation.}\label{fig:graph}
\end{figure}

\subsection{Direct Feature-based Model}

Defined in this way, we can focus on finding a feature mapping and regression function for recommendation. We can get $\mathbf{x}_u$, $\mathbf{x}_i$ and $\mathbf{x}_k$ for each user, item and feature by collaborative learning of the embeddings. In this stage, we try to build our model based on latent Dirichlet allocation. Because LDA model operates on documents, we should define the concept of the document in our model. Naturally, we can derive the document from the review text, thus, we define a review as a document. The corpus of LDA $\mathcal{D}$ includes all reviews in the dataset. We hope to extract and model user preferences, item profiles and feature embeddings from textual reviews in the data.

First, we can define the topic distribution for a user and an item, written as $\mathbf{x}_u$, $\mathbf{x}_i$. We use $k$ to stand for a feature, and $u$ and $i$ stand for user and item, respectively. We start with the likelihood function of LDA model. Because we define a document in the LDA model as a review $r$ that user $u$ gives to item $i$. So we can write the feature distribution of a document as $\mathbf{x}_{r,k} = P(k|u,i)$. Furthermore, we assume the relationship between user, item and feature as the probabilistic graph in Fig.1 , which means that a user chooses a feature and then chooses an item based on the feature. Based on the graph, we can know the user $u$ and item $i$ are conditionally independent given feature $k$. Because parameters $\mathbf{x}_{r,k}$ show the review properties in $K$ features, while $\mathbf{x}_{u,k}$ and $\mathbf{x}_{i,k}$ describe the $K$-feature distribution of a user and an item, so we can use the doc product between them to gain the ranking score of the user on the item over a particular set of feature(s) to generate the recommendation results. 

Based on the above derivation, we can design the recommendation score function as:
\begin{equation}
\begin{aligned}
\label{eq:model}
\hat{s}_{u,i} = \sum_{k}\mathbf{x}_{u,k}^T \mathbf{x}_{i,k}
\end{aligned}
\end{equation}
which summerizes the ranking scores over $K$ features. With this function, we can know the user's overall preference on an item based on all features in the feature space. 

\subsection{Similarity between Topic-based and Factor-based Models}

We want to bridge some connection between the latent topic in topic models and the latent vectors in matrix factorization models. The parameters $\mathbf{x}_{u,k}$ and $\mathbf{x}_{i,k}$ show the user's and item's properties in $K$ features, which can be considered in both topic models and latent factor models. Both types of models are trying to learn the user and item representation vectors in some feature space. As a result, the meaning of the user or item representation vectors are very similar for these two types of models.

\section{Cold-start Recommendation}\label{sec:model-cold}

The existing model faces serious cold-start problem. When we have few information about a new user, we cannot train meaningful $\mathbf{x}_{u}$ and $\mathbf{x}_{i}$ vectors to describe the user and item. This situation is especially important in hotel recommendation because people usually travel to each place for only one or two times, especially for entertainment trips. Usually, the system is required to make recommendation for a user in a new place, which is the first time for the user to go to the place, as a result, there is little personalized information from that place that can be used to make recommendation for the user. 

To alleviate the problem, the user's reviews in the user's behavior from other places can be used. The reason is that though the hotels can be very different in different places, the key features of the hotels can be similar. For example, the frequently used features to describe a hotel include price, service, environment, room size, parking lot, etc. These features frequently appear in user reviews, as a result, by learning the user and item latent vectors on the feature-level as shown in equation \eqref{eq:model}, the model helps to reduce the cold-start problem in recommendation.

Furthermore, we need to capture the different user behaviors in different places, e.g., some users prefer their own traditional preference on features no matter in which place, while others may care about different features in different places, such as a user may care about service on business trips, while they may care more about the environment on entertainment trips. Our goal is to design a model that can describe the user's preference on different features for different places. According to these considerations, we get the improved prediction function by adding the aggregated feature quality for each place $p$,
\begin{equation}
\label{eq:cold-model}
\hat{s}_{u,i} = \sum_{k}\mathbf{x}_{u,k}^T \mathbf{x}_{i,k} + \mathbf{x}_{u,k}^T \mathbf{x}_{p,k}
\end{equation}
where the $\mathbf{x}_{p,k}$ describes the place $p$'s performance on feature $k$ according to the reivews of all hotels located in place $p$. We can use this function to calculate the ranking scores for each user-item pair.

\section{Model Optimization}
\label{sec:optimization}

For fitting the model proposed in above sections, we develop a loss function to minimize. For simplicity, we use the standard sum square loss:
\begin{equation}
\begin{aligned}
\mathcal{L} = \sum_{s_{u,i}\in R}\big(s_{u,i} - \hat{s}_{u,i}\big)^{2} + \lambda \Omega(\Theta)
\end{aligned}
\end{equation}
where $\mathcal{L}$ is the loss, $R$ is the set of training data points, and $\lambda$ is regularization coefficient for the model parameters. Our goal is to find a set of parameters to minimize the loss function of our dataset.
\begin{equation}
\mathop{argmin}\limits_{\Theta} \mathcal{L}(\Theta,R)
\end{equation}

\section{Case Study}\label{sec:case}
Before introducing the experimental results of our model, we conduct an initial study of the hotel features. The data is from Tripadvisor, a major trip planning and hotel recommendation website.
\begin{figure}[t]
\centering
\includegraphics[width=0.9\linewidth]{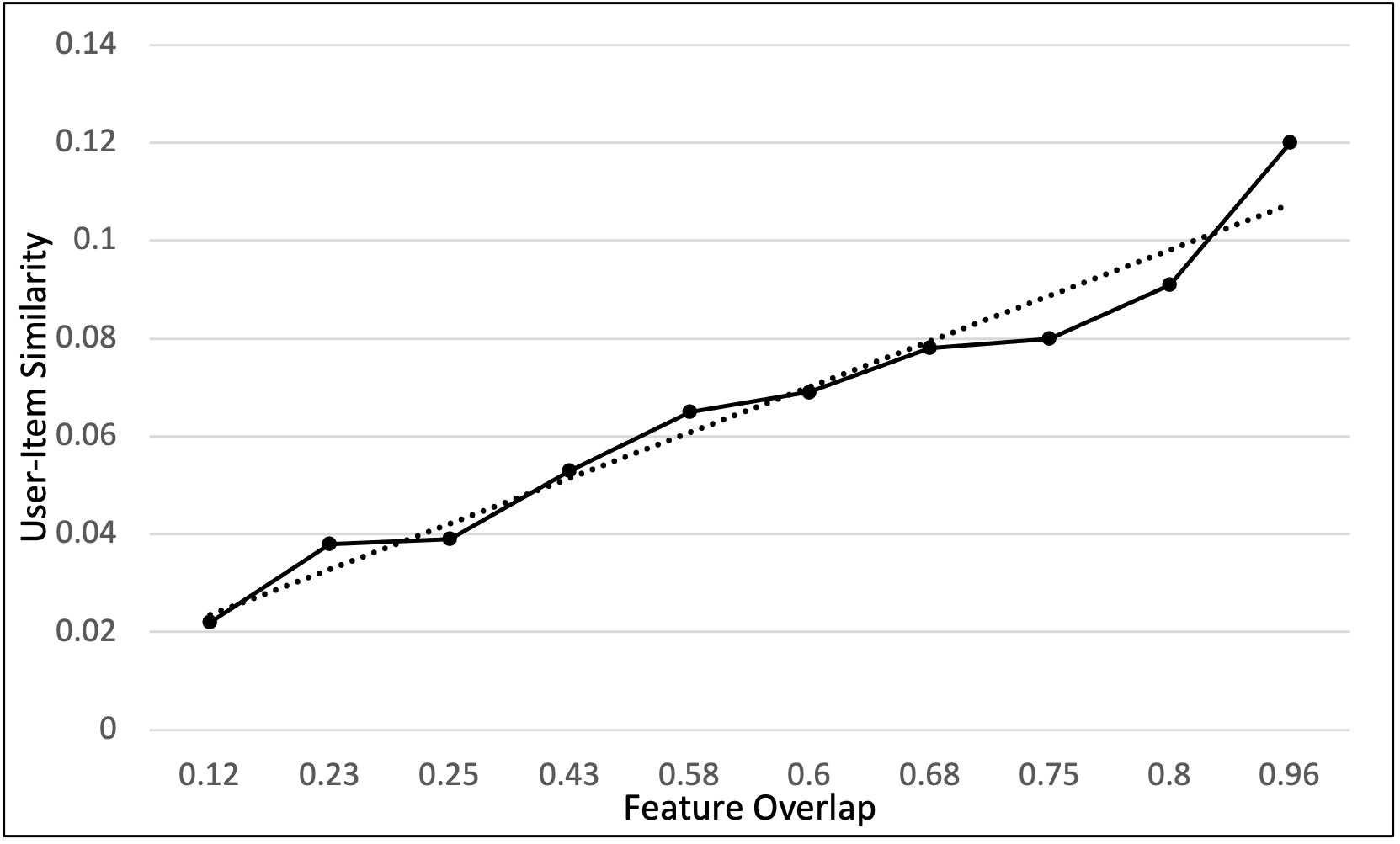}
\caption{The correlation between user-item similarity and average feature overlap.}\label{fig.2}
\end{figure}

\subsection{Feature Analysis}
In this stage, we investigate different features of hotels. Fig.\ref{fig.2} shows the relationship between the user-item similarity and their feature overlap. From this result, we can see that the feature overlap is a good signal to predict the similarity between users and items and thus to make recommendations, which implies that it is reasonable to use features as a good source of information to make hotel recommendations. In traditional models, which did not consider the regional property of features, we will face serious cold-start problems. For instance, when a user comes to a new place, since all of the user's previous interactions belong to other places, it will be difficult to provide recommendations for the user. However, the feature overlap will help us to make recommendations even in new places because hotels even in different places will share some common features.

\begin{table}[t]
\vspace{-15pt}
\caption{Top three features of some sampled hotels. Each column corresponds to a particular hotel.}
\centering
\begin{tabular}
	{c c c c} \hline
    location & location & service & staff\\
    service & price & staff & airport\\
    room & good & spacious & decoration\\\hline
\end{tabular}\label{tab:features}
\end{table}

\subsection{Topical Feature Examples}
To analyze the topical features in the review text, we use the LDA model to find the topic words. Table \ref{tab:features} shows some features for several sampled hotels. We can see that users cares about some frequently considered features such as price and location, while there are also some words such as ``good'' that are not features due to errors in feature extraction.

\section{Experiments}\label{sec:experiment}
In this section, we conduct experiments to study the performance of our model. We study the performance of the model on rating prediction and top-N recommendation tasks.
In the following, we will give experiment results on two different metrics, RMSE and Precision. We compare with the following models:
\begin{itemize}
    \item MF: The matrix factorization model, which learns user and item latent vectors based on user-item rating matrix for recommendation.
    \item BPR-MF: The frequently used Bayesian Personalized Ranking model for recommendation, we use matrix factorization as the base model for prediction.
    \item HFT: The Hidden Factor and Hidden Topic model for recommendation, which combines factorization and topic model for recommendation.
\end{itemize}

\begin{figure}[t]
\centering
	\subfigure[RMSE]{\label{fig:rmse}{
		\includegraphics*[viewport=0mm 0mm 192mm 143mm, scale=0.2]{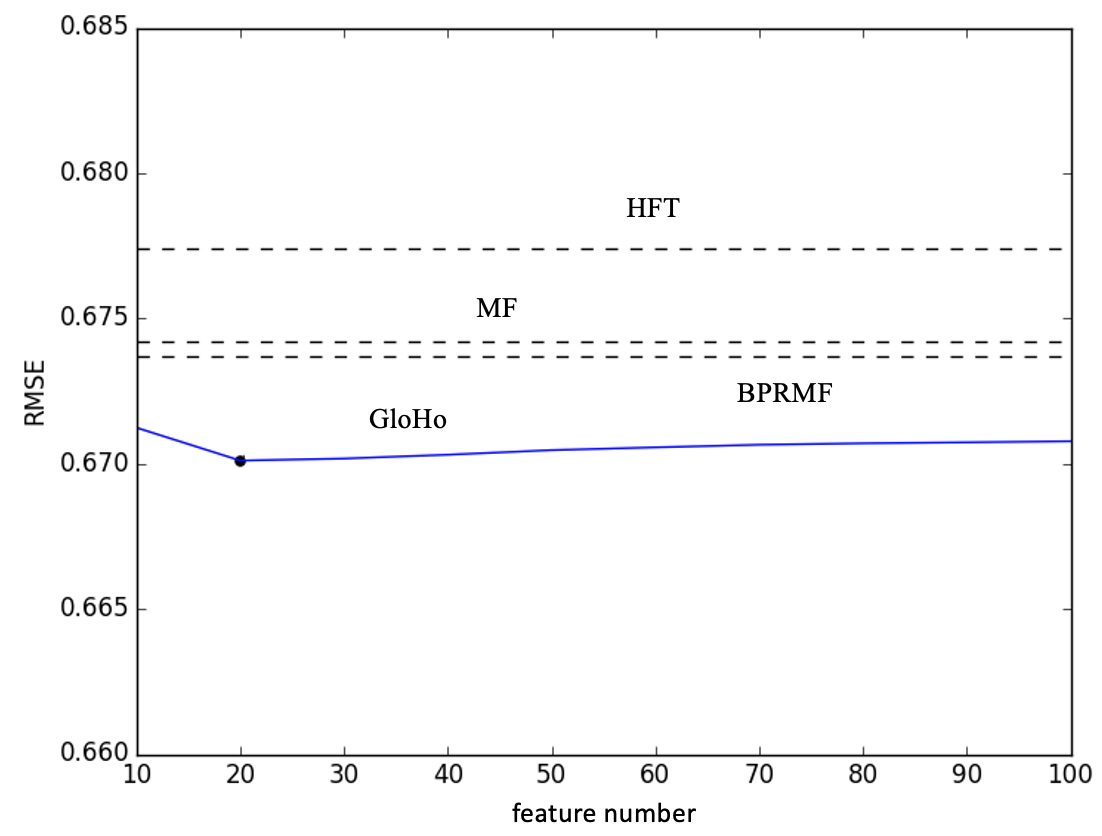}}}
\hspace{-10pt}
	\subfigure[Precision@5]{\label{fig:prec}{
		\includegraphics*[viewport=0mm 0mm 192mm 143mm, scale=0.2]{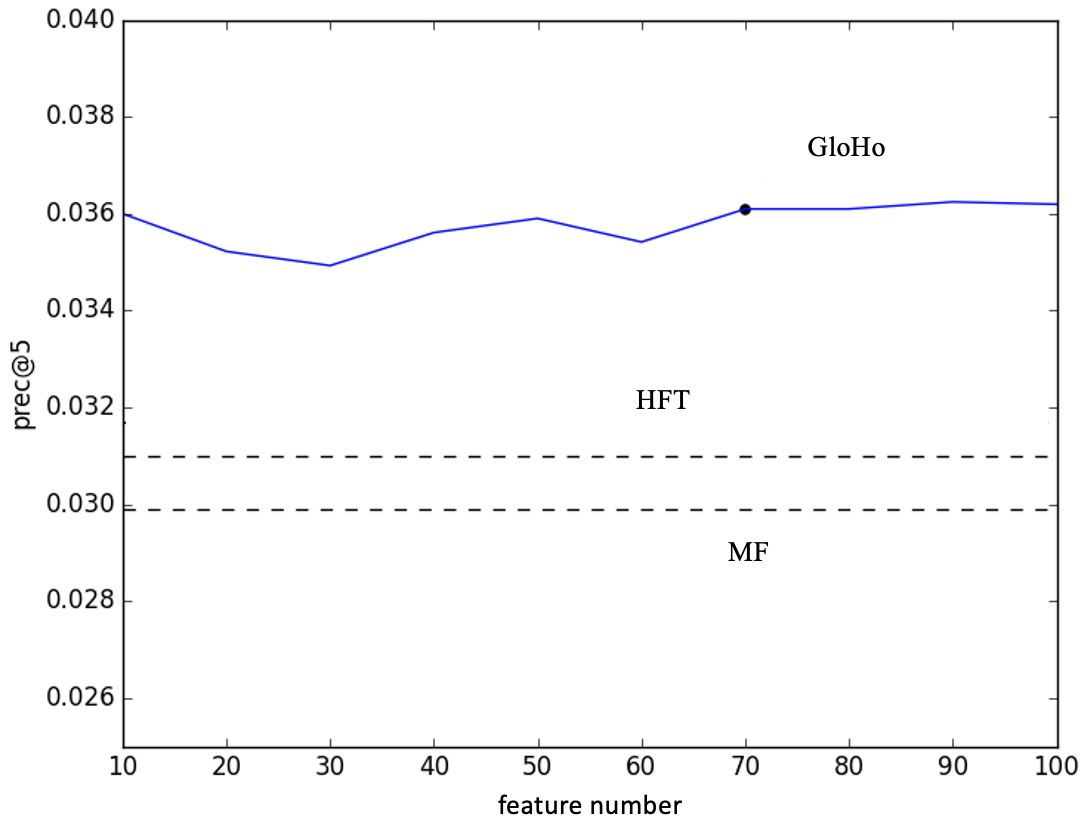}}}
\vspace{-2ex}
\caption{Performance on RMSE and Precision under different number of features.}
\label{fig:performance}
\end{figure}

\subsection{Rating Prediction Performance}
In this section, we study the model performance in approximating the ratings. We first evaluate our model in the traditional recommender situation in 5-fold validation. The experimental results of RMSE under different numbers of features are shown in Figure \ref{fig:rmse}. The standard deviations in 5-fold cross-validation of each baseline algorithm is $\le$ 0.005. We see that GloHoRec is better than the baselines. We can know that integrate the feature information is helpful, and this further validates the underlying intuition of adopting review features for item modeling in recommender systems.

\subsection{Ranking Performance}
The result on top-N recommendation performance is shown in Figure \ref{fig:prec}. The results are evaluated on precision at top 5 recommendation. We can see that the GloHoRec model is still better than baselines. Besides, we can see that the result of $Prec@5$ slowly grows with the increasing of feature numbers. Compared to the result of RMSE, we can see that user preference on choosing a hotel or not is more influenced by the features than the ratings on the hotel.

\section{Conclusion and Future Work}\label{sec:conclusion}

In this paper, we consider using item features for the problem of hotel recommendation. To solve the problem, we propose a Global and Local feature model for Hotel Recommendation (GloHoRec), which makes use of the features for rating prediction and recommendation. Our analysis of regional features shows that different hotels have different features that are cared about by users, which influence the user preferences on hotels. By further taking into account the place of hotels, our model is able to provide recommendations with regard to both the users' historical preferences and the characteristics of the hotels in a personalized way. Experiments on rating prediction and top-N recommendation tasks show the improved recommendation performance of our model. 

Our model is not restricted to the hotel recommendation scenario, but also applicable to other recommendation problems, which we will consider in the future. Furthermore, we can also consider state-of-the-art neural language models such as BERT to learn the feature embeddings so as to get the user and item latent vectors. This will help to better capture the semantic information in the reviews for better recommendation, which we will consider in the future.

\bibliographystyle{IEEEtran}
\balance
\bibliography{paper}

\end{document}